\RequirePackage{fix-cm}
\documentclass[smallextended]{svjour3}       
\smartqed  
\usepackage{graphicx}

\journalname{Experiments in Fluids}
\usepackage{natbib}
\begin{document}

\title{Lagrangian tracking of colliding droplets 
\thanks{This material is based upon work supported by the National Science Foundation under Grant No. 1605195.}
}

\subtitle{}

\author{RV Kearney         \and
        GP Bewley
}

\institute{Reece V. Kearney \at
              Cornell University
              124 Hoy Road, Ithaca NY 14850 \\
              Tel.: (845)-802-3001\\
              \email{rvk28@cornell.edu}
              \and
              Gregory P. Bewley \at
              Cornell University}

\date{Received: date / Accepted: date}

\maketitle

\begin{abstract}

We introduce a new Lagrangian particle tracking algorithm that tracks particles in three dimensions to separations between trajectories approaching contact. 
The algorithm also detects low Weber number binary collisions that result in coalescence as well as droplet break-up.
Particles are identified in two-dimensional high-resolution digital images by finding sets of circles to describe the edge of each body.  
This allows identification of particles that overlap in projection by over 80\% even for noisy images and without invoking additional temporal data.  
The algorithm builds trajectories from three-dimensional particle coordinates by minimizing a penalty function that is a weighted sum of deviations from the expected particle coordinates using information from four moments in time. 
This new hybrid algorithm is validated against synthetic data and found to perfectly reproduce more trajectories than other commonly used methods.  
Collisions are detected with 95\% accuracy for particles that move on average less than one tenth the distance to their nearest neighbor.  


\end{abstract}

\section{Introduction}
\label{intro}

The organization and coalescence rate of inertial droplets in turbulent environments is of great importance 
in a number of natural and industrial systems.  
In combustors, the size distribution of droplets affects the rate of evaporation of fuel and thus combustion efficiency, as in \citet{Betelin2012}.  
In clouds, droplets must grow large enough in order to fall as rain.  
Inability to predict droplet growth across the so-called ``size gap'' \citep[][]{Xue2008, Grabowski2013}, 
the regime of droplet sizes from about 10 microns to 100 microns in which droplets grow faster than condensation and differential gravitational settling can explain, is one of the greatest uncertainties in climate and weather modeling \citep[][]{Devenish2012}.  

Inertial droplets and their near-collision and coalescence have been studied extensively.  
Analytical and numerical work predicted the growth rate of droplet populations.  
\citet{Smoluchowski1916} introduces a stochastic collection equation that includes the notion of a collision kernel, 
a measure of the collision rate.  
\citet{Saffman1956} shows how the collision kernel can be expressed for droplets in turbulence.  
\citet{Maxey} describes how particles settling in turbulence cluster into the straining regions of the flow.
Direct numerical simulations (DNS) of inertial particles have been instrumental in showing that turbulence induces clustering as a function of the particle inertia and Reynolds number of the flow and can enhance collision rates \citep[][]{Reade2000, Chen2006, Salazar2008, Wang2000, Ayala2008, Ireland2016, Ireland2016a, Olivieri2014},
and many of these findings have been successfully validated in experiments \citep[][]{Sumbekova2016, Obligado2014}.
Combined experimentation, simulation, and modeling have also illuminated the behavior of inertial particles sedimenting under the influence of gravity \citep[][]{Lapp2012, Aliseda2002, Good2014, Salazar2008, Kawanisi2008, Ireland2012, Siewert2014} 

Experiments measuring collisions to complement these results are lacking.
Three-dimensional Lagrangian particle tracking (3D LPT) is used routinely in research to measure fluid motions \citep[][]{Xue2008, Gulan2012, Holzner2008, Hoyer2005, Virant1997, Ott2000},
and the motions of non-tracer particles in fluid \citep[][]{Lapp2012, Aliseda2002, Bewley2013, Salazar2008},
but these works do not give direct measurements of collisions from 3D LPT.

This deficit in experimental data is partly due to the difficultly of measuring the motions of particles when they are close together \citep[][]{Ouellette2006}.
To avoid this difficulty, researchers have successfully conducted experiments to characterize the preferential concentration and collision rate of populations of droplets by other means \citep[][]{Bateson2012, Duru2007}.
Researchers have also recorded the coalescence outcome of collisions between two droplets given the Weber number and impact geometry \citep[][]{Qian1997}, 
but these studies do not explain how the dynamics of droplets are affected by background flow 
or provide a statistical description of collision rates given ambient flow conditions.  

\citet{Schanz2016} developed a family of new tracking algorithms known as Shake the Box (STB).  
Whereas older methods search for particles in camera images independent of the tracking phase, 
STB methods use the time history of each measured trajectory to inform the search for particles in subsequent images. 
The predicted location of each particle is projected onto a camera image and ``shaken'' to optimize the difference between the measured pixel intensity and the anticipated effect of the particle that has moved to that location.
This allows trajectories to come much closer together.  
The emphasis of this algorithm has been on increasing the maximum particle seeding density that can be tracked, 
and to our knowledge, the methods have not been applied explicitly to study the motions of colliding particles.  

We believe that the only experiment that directly measured collisions of droplets in a turbulent environment is reported in \citet{Bordas2013}. 
The software used in this experiment, however, required user intervention to discern whether or not collisions had occurred, 
and the uncertainty in the result is not clear. 
Ultimately, to understand the near-contact dynamics of particles, a new particle tracking tool is necessary.

\section{Methods}
\label{sec:1}

Particle tracking can be broken into three steps: 
the identification of particles in two dimensional images on each camera, 
stereoscopic reconstruction of three-dimensional coordinates from several simultaneous camera images, 
and the building up of 3D trajectories in time to determine what particle is the same from frame to frame.  
In this paper we describe advancements in the first and third steps of this process.
We do not discuss direct improvements to the second step of particle tracking, though the process of stereoscopic reconstruction of particle coordinates can benefit from more precise two-dimensional particle identification and particle size measurement.
Additionally, erroneously constructed three-dimensional particles or ``ghost particles'' that result from errors in the reconstruction process can be more easily identified with better particle tracking.

\subsection{Particle Identification}
\label{sec:2}

In typical particle image velocimetry (PIV) and LPT applications, particles are small; their diameters are on the order of a few pixels or less.  
These methods are often used in order to measure the motions of the flow,
so minimizing the particle size in order to make them better tracers of the flow is important.  
Because we are interested in inertial particles, we instead take advantage of a higher resolution regime of particle images 
wherein particles have tens of pixels on their circumference.  
This allows us to use the particle size as a consideration to assess whether or not a collision has occurred during the collision detection phase.

The primary difficulty of differentiating between particles at small separations in a two-dimensional image 
is identifying the distinct projections within a body as well as identifying their positions and sizes in pixel space.
Consider a spherical particle at some position in three-space $\vec{x_w}$.  
This particle is recorded by a camera and a two-dimensional projection is created on the sensor.  
The projection of the particle leaves a circular spot.  
When two or more particles are close together, 
individual spots can result from several overlapping circular projections.  
We will call this agglomerated bright spot a single body.    

The new method discussed here uses only the edge data from each body.  
Many methods exist for finding edges in digital images 
\citep[][]{Canny1983,Deriche1987,Sobel1968,Trujillo-Pino2013}.  
We used the following process: 

\begin{enumerate}

\item Apply a Gaussian blur to the image to reduce the effect of noise \citep[][]{Kobayashi1991}.
\item Calculate the gradient of pixel intensity of the image.
\item Find the pixel-accuracy boundary of the body using the coordinates of pixels above some brightness threshold, $T$.
\item For each pixel-accuracy boundary point, interpolate on the grid in the direction of the gradient to find the subpixel location 
where intensity falls below threshold $T$.

\end{enumerate}

\begin{figure}
\includegraphics[width=\textwidth,height=\textheight,keepaspectratio]{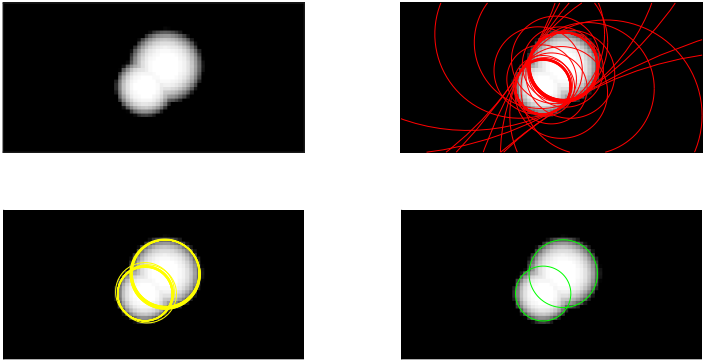}
\caption{{\it Top left:} An example of two synthetic particle images that are overlapping by 60\%
{\it Top right:} Circles fit by the Pratt method \citep[][]{VaughanPratt1987} to contiguous data points along the boundary of the body ($Q = 11$)
{\it Bottom left:} The circles that remain after those with centers outside the body were discarded
{\it Bottom right:} The results of averaging the properties of families of circles that have center coordinates within the same peak of the accumulator matrix}
\label{fig:1}       
\end{figure}

Given the edge coordinates of a body, the task is to find the set of circles that best describes the body.  
This is a combinatorially complex process, and so we must formulate a reasonable way to continue.  
The algorithm we propose, which we name ``Pratt-Walking'', is described pictorially in Figure \ref{fig:1} and proceeds as follows.  

\begin{enumerate}

\item Consider a set of edge data that consists of $N$ two-dimensional coordinates.  
\item A contiguous set of $Q<N$ edge data points are selected and a circle is fit to the data using the method of \citet{VaughanPratt1987}.
\item This process repeats for all $N$ sets of $Q$ contiguous data points, yielding $N$ circles (see \ref{fig:1}, \textit{Top right}).
\item Erroneous circles are discarded (circles with centers outside the bounding rectangle of the body and circles whose edges extend far outside the bounding rectangle of the body; see \ref{fig:1}, \textit{Bottom left}).
\item The remaining circles are binned according to their center positions with a bin-width of 1 pixel.
\item Contiguous sets of bins that represent a number of circles greater than some threshold $H$ are associated with populations of circles that have similar center coordinates. The properties of these circles are averaged to generate a reduced family of circles (see \ref{fig:1}, \textit{Bottom right}).
\item A residual is calculated that gives the average distance from the edge data and the nearest edge of a circle in the reduced family. If this residual is smaller than some threshold $G$, then the solution is accepted.

\end{enumerate}

\subsection{Particle Tracking}
\label{sec:3}

Given the coordinates of a set of particles in space and time, 
particle tracking refers to the process of identifying which particle is the same from frame to frame.  
Here, we refer to a linked set of coordinates in time as a trajectory.
When there are many particles during each of many time steps, 
constructing trajectories is an NP-hard multidimensional assignment problem \citep[][]{Veenman2003}.  
To avoid this computational difficulty, tracking algorithms use only a small subset of particle coordinates at a time \citep[][]{Ouellette2006}.  

When particles are far apart from one another relative to how far they travel, 
tracking their positions is easy since the number of physically realizable trajectories is small.
As the particle density increases or the particles move faster, holding all else constant, 
the number of physically reasonable choices for the next point in each trajectory increases.
Regardless of the tracking method chosen, tracking performance degrades as particles move faster relative to the distance between them \citep[][]{Ouellette2006}.
If measurements of collisions are to be taken, it is impossible to avoid the problem of particle coordinates coming arbitrarily close together, 
and so it is necessary to form a robust, quantitative metric to determine the best match from the identified particle coordinates in the next time step for each trajectory.
Collision detection will take place in a distinct step prior to the tracking process, as described in Section \ref{sec:4}.

Consider a trajectory $t_n$ that has been successfully tracked until position $\vec{x_{n,i}}$ at time $\tau_i$.  
At time $\tau_i$, $t_n$ also has some estimate for its velocity $\vec{v_{n,i}}$.  
The algorithm seeks to extend this trajectory into the next time step $\tau_{i+1}$ 
by calculating a penalty function associated with each set of particle coordinates that are known at time $\tau_{i+1}$.  
This penalty function consists of the weighted sum of three components: 

\begin{equation} \label{eq:1}
P = w_1 d_1 + w_2 d_2 + w_3 d_3
\end{equation}

Where the $w_i$ are weightings.  
The $d_m$ are defined for $t_n$ matching with particle coordinates at position $\vec{y_{j,i+1}}$: 

\begin{equation}
d_1 = |\vec{x_{n,i}} - \vec{y_{j,i+1}}|
\end{equation}
\begin{equation}
d_2 = |\vec{x_{n,i}}+\vec{v_{n,i}}d\tau  - \vec{y_{j,i+1}}|
\end{equation}
\begin{equation}
d_3 = |\vec{y_{j,i+1}}+\vec{v_{n,i+1}}d\tau  - \vec{y_{k,i+2}}|
\end{equation}

where $d\tau = \tau_{i+1} - \tau_{i}$. 
$\vec{y_{k,i+2}}$ is chosen from all possible coordinates of particles at time $\tau_{i+2}$ such that $d_3$ is minimized.

These three $d_m$ correspond to three tracking methods described in \citet{Ouellette2006}, so we call it a ``hybrid method.''
$d_1$ is the distance between the current trajectory's coordinates and the coordinates of a candidate particle at the next time step.
It is identical to the tracking cost of the ``Nearest Neighbor'' heuristic.
$d_2$ is the difference between the predicted and observed position of particle coordinates at the next time step.
It is identical to the tracking cost of the ``Minimum Acceleration'' method.
$d_3$ is the difference between the predicted and observed position of particle coordinates in two time steps.
It is identical to the tracking cost of the ``Four Frame: Best Estimate'' with an estimated acceleration of zero at all times.
Determining the optimal weights to track a given population of particles is not trivial, but we will show (in Section \ref{sec:8}) that even for unoptimized, reasonably chosen weights the hybrid method outperforms other methods.

We also apply a threshold on the maximum distance a particle can travel in one frame:

\begin{equation}
|\vec{x_{n,i}} - \vec{y_{j,i+1}}| < M
\end{equation}

Excluding matches based on the maximum distance a particle can travel yields a reduced set of potential matches. 
We use velocity estimates based on two-point, forward finite differences.
At the beginning of trajectories, it is not possible to formulate an estimate for the velocity using a finite difference, so we set $d_2 = M/2$.
For this reason, trajectories achieve substantially higher values of $P$, the penalty function, near their beginning compared with trajectories consisting of two or more points
The optimal pairs of trajectories and candidates are those that minimize the sum of $P$ across every trajectory, and they can be found with the Munkres Algorithm \citep[][]{Bourgeois1971}.  

\subsection{Collision Detection}
\label{sec:4}

Consider two spherical droplets of radius $r_1$ and $r_2$ that collide and coalesce.  
Neglecting fragmentation, the radius of the daughter droplet is $r_3 = (r_1^3+r_2^3)^{\frac{1}{3}}$.  
In addition, neglecting losses to heat and external forces, the momentum of the center of mass of the system must remain the same.  
These are the criteria we used to determine whether or not a collision occurred.

For each trajectory at time $\tau_i$, 
before the track is extended into frame $\tau_{i+1}$ as described in the previous section, 
the anticipated path of the trajectory is constructed.  
The algorithm then searches for intersections and near-intersections of these predicted paths.  
If one is found, the algorithm searches for a daughter at time $\tau_{i+1}$ at the location predicted by conservation of momentum.  
The algorithm also searches for the daughter at time $\tau_{i+2}$, again at the location predicted conserving momentum.  
If all these conditions are satisfied, then a collision is deemed to have taken place, and a link is formed between the two parent trajectories and the daughter trajectory. 
It is only necessary to check these criteria for trajectories that are closer than some distance $d < 2M+r_1+r_2$, where $r_1$ and $r_2$ are the radii of the particles.
Whereas previous tracking algorithms have produced particle tracks that have only one beginning and one end, 
this allows trajectories to converge so that a trajectory may have many beginnings, each associated with parents, 
and one end, associated with the ultimate daughter.
Droplet bursting can also be detected by feeding particle coordinate data in reverse-chronological order.
This method is suited for detecting collisions between liquid droplets at low Weber number.

\section{Results}
\label{sec:5}

\subsection{One-Particle Identification Test}
\label{sec:6}

\begin{figure}
\includegraphics[width=\textwidth,height=\textheight,keepaspectratio]{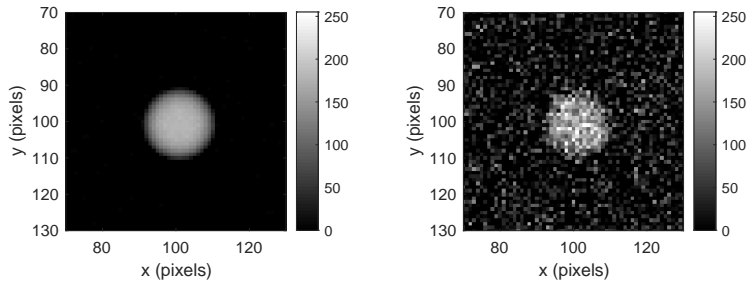}
\caption{
{\it Left:} Synthetic particle image in 1\% noise conditions.
{\it Right:} Synthetic particle image in 20\% noise conditions.}
\label{fig:7}
\end{figure}

We tested the performance of the algorithm using synthetic data.  
Synthetic images were constructed by generating random particle coordinates and finding the effect of each particle on the pixels of the image.  
A quartic brightness intensity profile was chosen such that the effect on a pixel due to a particle of radius $r$ located some distance $dr$ away from the particle center is given by  
\begin{equation}
I(dr) = I_{max} - (I_{max}-I_{min})\frac{dr^4}{r^4}
\end{equation}
$I_{max}$ was chosen to be 180 and $I_{min}$ was chosen to be 10, 
consistent with an 8-bit image with good contrast between particle and background.  
Zero-mean Gaussian noise was added to the image to test the robustness of the algorithm.
The standard deviation of the noise was set as a percentage of the maximum brightness.
Sample images of the synthetic particle at two different noise levels can be seen in Figure \ref{fig:7}.

\begin{figure}
\includegraphics[width=\textwidth,height=\textheight,keepaspectratio]{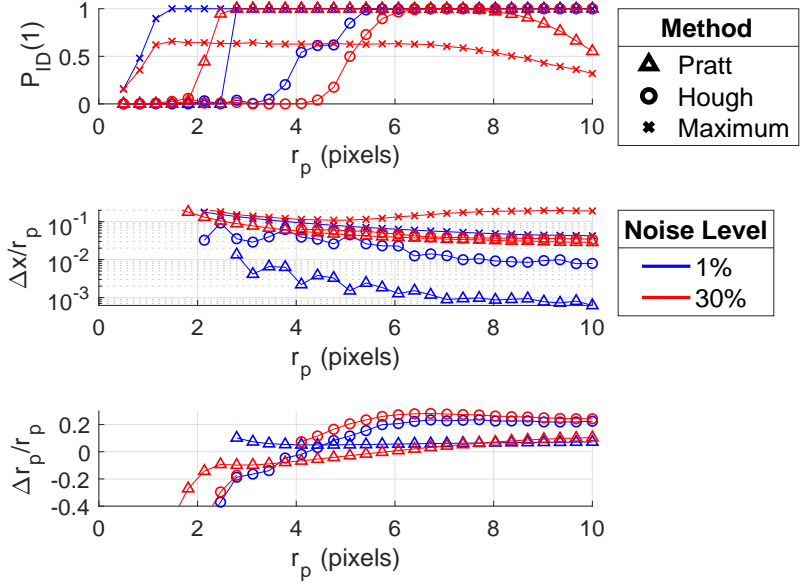}
\caption{
{\it Top:} The probability of correctly identifying a particle in a simulated image of an isolated particle for different noise levels and particle sizes.  
{\it Middle:} The average error in measuring the center location of the particle.  {\it Bottom:} The average error in measuring the particle radius. A negative value indicates an overestimate of particle size. The local maximum method does not produce an estimate of particle size.
{\it Legend:} Symbols correspond to the identification method used. Color corresponds to the standard deviation of the Gaussian noise added to the images. 
In low-noise conditions, the center-finding error for Pratt-Walking is about 20 times smaller than the local maximum method and about 10 times smaller than the Hough Transform method.}
\label{fig:2}       
\end{figure}

Figure \ref{fig:2} shows the results of the algorithm on an image containing a single particle.  
The results of the algorithm described here are compared to a circular Hough Transform as implemented in the MATLAB function \textit{imfindcircles}.  
We also compare against a simple method that searches for local maxima above threshold $T$ in the pixel brightness distribution after a Gaussian blur.
This method does not achieve sub-pixel precision and is intended as a benchmark naive and fast process
The results shown are averaged over 10,000 trials.  
In each trial, the synthetic particle image is generated in the center of a 200 by 200-pixel frame and offset by a random, 
sub-pixel amount in both the x and y-directions.  
Pixel intensities are bounded such that they can only achieve values between 0 and 255 (consistent with 8-bit images).  
We use the following values for the algorithm parameters (see Section \ref{sec:2} for definitions): $T = 55 ; Q = 11; H = 11; G =1$.  
The standard deviation of the Gaussian filter is $\sqrt{3}$ pixels.  

We define $P_{ID}(n)$ as the probability of the algorithm detecting that there are $n$ particles in the image.
The accuracy of finding the center of the particle is given by $\frac{dx}{r_p}$, where $dx$ is the distance between the measured particle center and its actual center and $r_p$ is the actual particle radius. 
The accuracy of size estimation is given by $\frac{dr_p}{r_p}$ where $dr_p$ is the difference between the actual particle radius and the measured particle radius.

Pratt-Walking identifies a wider range of particles of varying size with greater reliability and superior center-finding precision than either of the other methods considered at all noise levels. 
The Hough method is able to give better estimates for particle sizes than Pratt-Walking for a small regime of particle size (between 4 and 5 pixels in radius) at low noise conditions only. 
The local maximum method is also able to more reliably identify particles with radii smaller than 3 pixels than Pratt-Walking at the lowest noise conditions, but its error in measuring the particle center increases steeply in this regime.

Note that the radius of the smallest particle identifiable by Pratt-Walking decreases with increasing noise, which may be counterintuitive.  
This is because the method relies on data at the edge of the body, where the true pixel intensity should be close to zero.  
While the noise added is Gaussian with a mean value of zero, a pixel's minimum value saturates at zero, 
so the average pixel intensity imparted by the noise near the boundary is positive.  
This also leads to systematic overestimates in the size of the particle, as shown in the bottom of Figure \ref{fig:2}.  

Pratt-Walking performs worst in terms of particle identification, with poorer performance than the Hough Transform method for particles larger than 8 pixels in radius for moderate noise conditions (see \ref{fig:2}: \textit{Top}).
This is a result of the choice of $Q$ and $H$.  
Increasing $Q$ and $H$ would give better estimates for all three of the metrics considered in Figure \ref{fig:2} for large particles, but would degrade performance for small particles. 
In general, care must be taken in selecting the values for $Q$, $H$, and $T$ based on the size and brightness of the particles being tracked.  

Both the Hough Transform and Pratt-Walking exhibit radius error that changes as a function of the particle size and noise conditions.  
The near-constant error of Pratt-Walking at low-noise conditions is due to the choice of $T$, 
which was selected to be high enough to ignore ambient noise throughout the image, 
but this causes an underestimate of particle size.  
As discussed above, Pratt-Walking overestimates the size of small particles because noise energizes the intensity of bodies at their border.  
All of these errors can be easily corrected post-measurement with calibration on particles of known size.  
Note that the local maximum method does not produce an estimate for the particle size.  
When there is one local maximum in pixel intensity in a body, 
a natural estimate for the particle size is the square root of the area of the body above threshold $T$, 
but when multiple maxima exist, 
additional segmentation methods are necessary to produce a size estimate, 
and we do not consider such methods here.  

\subsection{Two-Particle Identification Test}
\label{sec:7}

Next, the algorithm was tested on synthetic data that contains the projection of two particles that overlap, a frequent occurrence in densely seeded flows.  
The overlap is defined as 
\begin{equation}
Ov = \frac{r_1+r_2-d}{2\min{(r_1,r_2)}}, 
\end{equation}
where $d$ is the center-to-center distance between the two particles, and $r_1$ and $r_2$ are the radii.  
The overlap is defined such that it is equal to 0 when the particles are just touching edges with their centers entirely outside one another 
and 1 when one particle completely obscures the other.  
$r_2$ is held fixed at 10 pixels while $r_1$ is varied so that different radius ratios can be achieved. 
 
\begin{figure}
\includegraphics[width=\textwidth,height=\textheight,keepaspectratio]{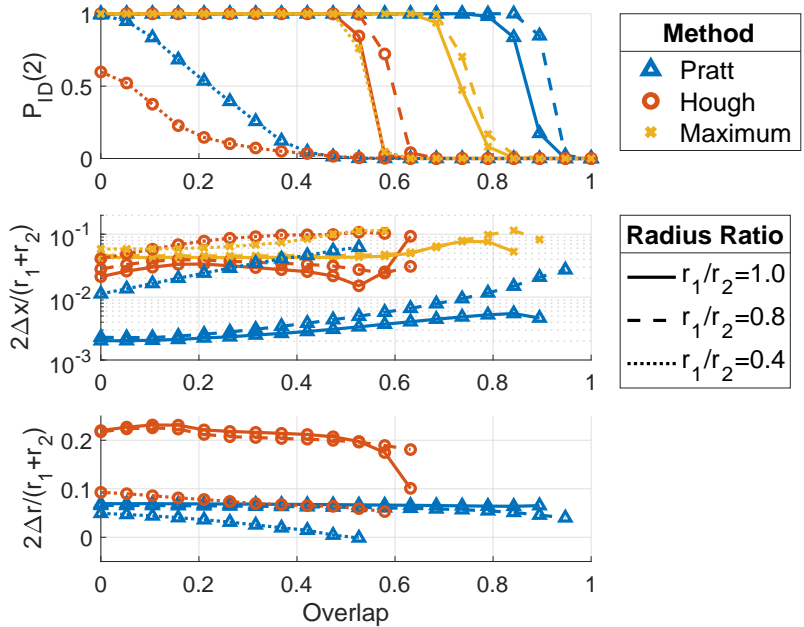}
\caption{Low noise (1\%) cases.  
{\it Top:} The probability of correctly identifying two particles in a simulated image of two overlapping particles.
{\it Middle:} The average error in finding the center locations of the particles.
{\it Bottom:} The average error in finding the particle sizes. The local maximum method does not produce an estimate of particle size.
{\it Legend} Symbols and color correspond to the identification method used. Line type corresponds to the ratio of the radii of the two particle images.
Pratt-Walking is able to able to identify particle pair images at overlap at least 40\% higher than the Hough Transform method.
The magnitude of the center-finding error for Pratt-Walking is at worst the same size as the error for the other two methods.
At best, the center-finding error for Pratt-Walking is 15 times smaller than the error of the Hough Transform method and 20 times smaller than the local maximum method.}
\label{fig:3}       
\end{figure}

The results for this test are shown in Figure \ref{fig:3} again compared to a Hough Transform and the local maximum methods.  
The results for each set of conditions were averaged over 10,000 trials.  
Noise was fixed at 1\% of the maximum pixel intensity.  

Here, the center error was the mean error averaged between both particles.  
The radius error was averaged in the same way.  
We used the following values for the algorithm parameters: $T = 55; Q = 8; H = 8; G =2$.
The standard deviation of the Gaussian filter was $\sqrt{3}$ pixels.  

Pratt-Walking identifies overlapping particle pairs with more reliability and has better center-finding accuracy at all values of overlap and radius ratios considered, except compared to the local maximum method at $\frac{r_1}{r_2} = 0.4$. 
Even in this case, while the reliability of Pratt-Walking is lower, it still maintains superior center-finding accuracy. 
The failure of Pratt-Walking for small radius ratios is due to insufficient information on the border of the body that is associated with the smaller particle. 
In general, Pratt-Walking requires $Q+H > N_p$ where $N_p$ is the number of edge data points associated with a single particle.
For images of small particles with very little noise, the local maximum method works best, and indeed this method still has a place as a useful tool for characterization of flows seeded with tracer particles at low number densities.
It is less suited for studies that require precise estimation of particle location and size.

The radius error is positive for Pratt-Walking and the Hough Transform method at all values of overlap except at high overlap of very unlike sized particles, indicating a consistent underestimate of size. 
The magnitude of the error is smaller for Pratt-Walking than the Hough Transform for each radius ratio. 

\begin{figure}
\includegraphics[width=\textwidth,height=\textheight,keepaspectratio]{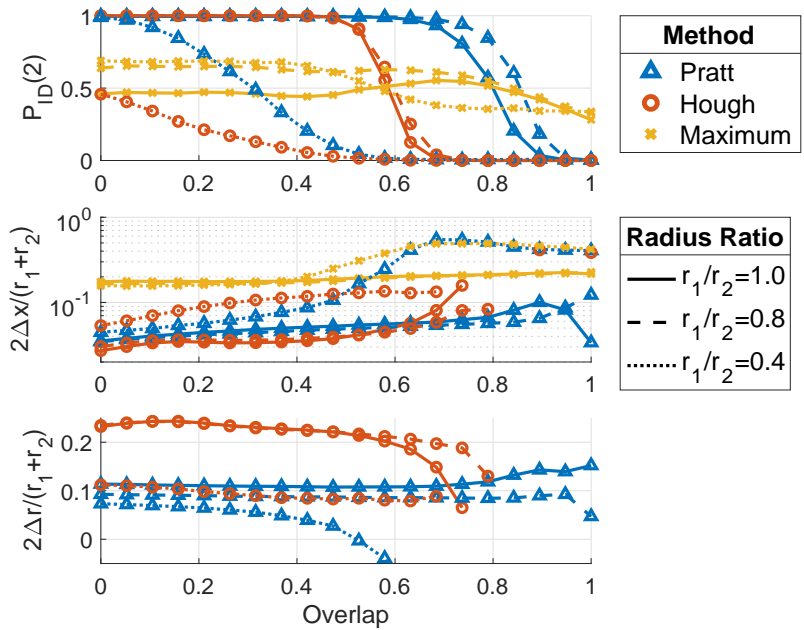}
\caption{Moderate noise (20\%) cases.  
{\it Top, Middle and Bottom:} as in Figure \ref{fig:3}.
The accuracy in center-finding and size-estimation for all three methods is decreased by the addition of noise.
Pratt-Walking is able to able to identify particle pair images at overlap about 30\% higher than the Hough Transform method.
The local maximum method does not reliably identify particle pair images at any overlap at this noise level.
The magnitude of the center-finding error for Pratt-Walking is about the same as the Hough Transform method and between 1.5 and 5 times smaller than the local maximum method.}
\label{fig:4}       
\end{figure}

Pratt-Walking, the Hough Transform, and the local maximum methods are also compared at moderate noise levels (20\% max pixel intensity). The results are shown in Figure \ref{fig:4}.

Noise does not substantially degrade the probability of identifying the correct number of particles for Pratt-Walking for $Ov < 0.6$, though it does dramatically harm the performance of the local maximum method, pushing the probability of correct identification below $0.8$ for all radius ratios at all values of overlap.
The Hough Transform method also suffers a substantial reduction in identification performance for $\frac{r_1}{r_2} = 0.4$.

The addition of moderate noise also causes significant reduction in performance of center-finding for all three methods. 
The accuracy in center-finding of the Hough method and Pratt-Walking becomes similar for $Ov < 0.5$, though Pratt-Walking maintains functionality through higher overlap values. 
Neither Pratt-Walking nor the Hough method exhibit errors over 0.1 for regimes in which the algorithm correctly identifies particles more than half the time. 
The local maximum method exhibits errors above 0.1 for all values of the overlap.

Noise also damages estimates for the particle sizes for the Hough method and Pratt-Walking, pushing radius estimates farther from 0, especially at high overlap. 
Errors associated with Pratt-Walking remain below 0.2 in absolute value for all overlap values and all radius ratios.

The local maximum method took the least time to process, and the Pratt-Walking and the Hough methods took about 2.5 and 3.8 times as much time as the local maximum method, respectively.

\subsection{Particle Tracking}
\label{sec:8}

In order to quantify the performance of our new hybrid particle-tracking algorithm, we tested it on synthetic data as in \citet[][]{Ouellette2006}.
The data were generated by a particle-laden turbulent flow solver called HiPPSTR and developed by \citet[][]{Ireland2013}.
For an isotropic turbulence at an average $R_\lambda = 52$ in a periodic cubic domain of length $2\pi$, statistical stationarity is maintained through deterministic forcing.
6400 tracer particles were present within the volume at all times.
Tracer particles have no inertia, and so have a Stokes number $St = \tau_p/\tau_\eta = 0$ where $\tau_p$ is the relaxation time of the particle and $\tau_\eta$ is the Kolmogorov time scale.

A measure for the difficulty of tracking a population of particles is given by $\xi$ which is defined as 
\begin{equation}
\xi = \frac{\delta x}{\delta d}, 
\end{equation}
where $\delta x$ is the average displacement of a particle from one frame to the next 
and $\delta d$ is the average separation between a particle and its nearest neighbor.

The quality of tracking can be measured by $E_{track}$, which is defined as 
\begin{equation}
E_{track} = \frac{T_{imperfect}}{T_{total}}, 
\end{equation}
where $T_{imperfect}$ is the number of tracks the algorithm failed to produce perfectly, 
and $T_{total}$ is the total number of tracks in the data set.  
A perfectly tracked trajectory must begin at the same time as an actual trajectory but need not end at the same time.  
This way, an initial break in a trajectory is not penalized more than subsequent breaks.  
Additionally, a perfectly tracked trajectory must contain no spurious points.  
This measure is identical to the one considered in \citet{Ouellette2006}.  

$\xi$ is altered by under-sampling the particle positions so that the average displacement of particles seen by the tracking algorithm is increased, resulting in trajectories that span the same spatial distance but have less information along that space. 
When a particle passes the periodic boundary and returns on the opposite side of the simulation volume, we consider it to be a new trajectory for purposes of calculating $\xi$ and $E_{track}$. 
This results in about 86,000 unique trajectories.

We use a simple finite difference to estimate all particle velocities.  
Rather than using the Munkres algorithm to resolve conflicts between trajectories, we use a simple first-come first-serve heuristic where the candidate with the better (smaller) score for the penalty function obtains the next point in the trajectory.  
We use the following weights in calculating the penalty function (see Equation \ref{eq:1}): 
$w_1 = 1; w_2 =5; w_3 = 4$.  
We set $M = 0.009f_s$ where $f_s$ is the sampling frequency.  

\begin{figure}
\includegraphics[width=\textwidth,height=\textheight,keepaspectratio]{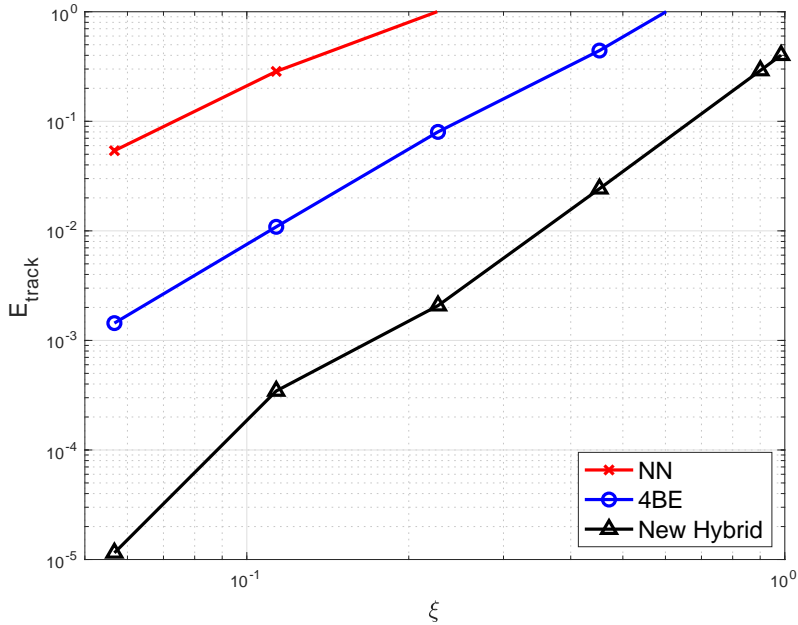}
\caption{We measured the performance of particle tracking methods 
with the tracking error, $E_{track}$, which is the fraction of imperfect tracks in the dataset, 
and which increases as a function of $\xi$, 
which measures how far particles travel relative to their distances from each other.  
We compared our hybrid method to select algorithms presented in \citet{Ouellette2006}, 
of which the Nearest Neighbor (NN) algorithm was the worst-performing 
and the 4 Frame: Best Estimate (4BE) method was the best-performing.
}
\label{fig:5}       
\end{figure}

Figure~\ref{fig:5} shows the mean values of $E_{track}$ for the hybrid algorithm compared to the best-performing and worst-performing methods from \citet{Ouellette2006}; 
these are respectively the Nearest Neighbor (NN) and 4 Frame: Best Estimate (4BE) heuristics. 
Note that our implementation of 4BE extended all trajectories that already contained two or more sets of coordinates before extending new trajectories, which enhances the performance of the method.
Our results for NN and 4BE differ from those in \citet{Ouellette2006} because our methodology for altering and computing $\xi$ differs.

The hybrid method achieves lower values of $E_{track}$ than the NN or 4BE methods for all $\xi$. 
The NN heuristic constructs imperfect trajectories whenever the separation distance between two trajectories is smaller than the distance traveled between time steps. 
This kind of event increases in frequency with increasing $\xi$.

The hybrid method outperforms the 4BE heuristic in two ways: by better initiating trajectories and by better continuing longer trajectories. 
For new trajectories, 4BE defaults to the NN heuristic, the worst-performing algorithm. 
In contrast, the hybrid method uses three frames of data to initiate trajectories, as discussed in Section 2.2.

We ran our tracking tests while enforcing the same perfect initiation for each of the tracking algorithms, so that every trajectory began with two correct sets of coordinates. 
As expected, every algorithm performed better. 
The difference in $E_{track}$ between the perfect-initiation test and the standard-initiation test gives the initiation error, and the final value of $E_{track}$ gives the error due imperfect continuation of longer trajectories. 
At tracking difficulty $\xi=0.45$, $E_{track}$ for 4BE was reduced by 0.15 to a final value of 0.29. 
$E_{track}$ for the hybrid method was reduced by 0.011 to a final value of 0.014. 
In other words, we found that even with perfect initiation, the hybrid method outperforms the 4BE method by an order of magnitude. 
The proposed hybrid method performs better in part because it incorporates the best features of each method, while excluding the large uncertainty inherent in acceleration estimates used by 4BE.

\subsection{Collision Detection}
\label{sec:9}

Similar to $E_{track}$, we define metrics for the quality of collision detection 
\begin{equation}
C_{g} = \frac{N_{g}}{N_{total}}, 
\end{equation}
\begin{equation}
C_{b} = \frac{N_{b}}{N_{total}}, 
\end{equation}
where $N_{total}$ is the total number of collisions that occur in the subdomain, 
$N_{g}$ is the number of collisions correctly identified, 
and $N_{b}$ is the number of false collisions identified 
(that is, the algorithm declares a collision occurred when in fact none did).  

\begin{figure}
\includegraphics[width=\textwidth,height=\textheight,keepaspectratio]{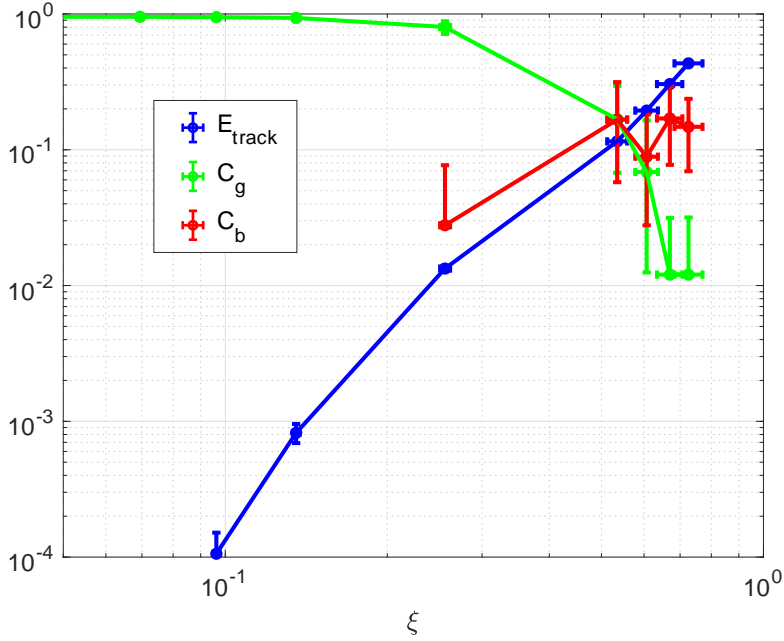}
\caption{We applied our algorithms to synthetic data to quantify the increase in tracking quality compared to other methods 
and to measure the accuracy in collision identification.  
The bars show 95\% confidence intervals for the mean of each quantity. 
When $\xi \approx 0.5$ the number of correctly identified collisions, which is measured by $C_{g}$, 
is equal to the number of false collision detections, measured by $C_{b}$.  
$E_{track}$ and $\xi$ are reduced here compared to Figure \ref{fig:5} because the decomposition of the domain into 27 subdomains artificially lowers $\xi$ without changing the motion of the synthetic particles.}
\label{fig:6}       
\end{figure}

The performance of the algorithm on DNS data with particle collisions is shown in Figure \ref{fig:6}.
HiPPSTR was modified to incorporate a collision algorithm \citep[][]{LiSingHow2020} and simulate coalescence assuming a unit collision efficiency, meaning that every particle crossing trajectory results in coalescence, whereby mass and momentum (but not energy) are conserved.
6400 inertial particles of identical size were allowed to reach statistical equilibrium over five large-eddy turnover times before the collision algorithm was activated, whereupon coalescence events were allowed to occur.
The particles also had an initial Reynolds number $Re_p = \frac{r_0u_\eta}{\nu} = 0.16$ where $r_0$ is the initial particle radius before any collisions took place, $u_\eta$ is the Kolmogorov velocity scale, and $\nu$ is the kinematic viscosity of the carrier fluid.
At the end of the simulation, 6308 particles remained.
The particles all had an initial radius of 0.0042 and $St = 0.1$.
To correctly identify a collision, we require that the collision occur at the correct time and location.  
Because coalescence is instantaneous in the simulation, collisions always occur between time steps, 
so we only require that the collision time be identified to within the interval between the preceding and following time step.  
We require a spatial accuracy of no more than $0.1M$.  
We break the domain into 27 subdomains of equal size without overlap in order to decrease the computational expense, 
which goes as the square of the number of contemporaneous trajectories.  
Additionally, we consider each daughter particle that results from a collision to be the start of a new trajectory for purposes of calculating $E_{track}$. 
This results in several hundred trajectories per subdomain over the course of the simulation. 
We perform a bootstrap process with 10,000 iterations on the 27 subdomains at each sampling rate to find the mean and 95\% confidence intervals of the mean at different values of tracking difficulty.
We use the following weights in calculating the penalty function: 
$w_1 = 1; w_2 =5; w_3 = 4$.  
We set $M = 0.008f_s$.

For values of $\xi$ smaller than 0.2, collisions are found nearly perfectly. 
No false collisions are detected in any case, and $C_{g}$ remains nearly constant at 0.95. 
As $\xi$ increases, $C_{g}$ decreases smoothly to about 0.01 at $\xi = 1$. 
When $\xi$ is small, the few errors in collision detection lead to a corresponding, small but finite tracking error. 

Note that the values of $E_{track}$ here are somewhat lower at the same value of $\xi$ than in the previous simulation with tracer particles. 
This is a consequence of subdividing the domain, which has two effects: less information is available near the edge of each subdomain, resulting in a degradation of $E_{track}$, and the space outside a subdomain is regarded as empty, causing a decrease in $\xi$.

These data suggest that detecting collisions imposes a more stringent requirement on the tracking difficulty than does simply tracking particle motion. 
At $\xi = 0.4$, only 2\% of the measured trajectories are imperfect, but more than half of the collisions are missed.

\section{Conclusions}
\label{sec:10}

We introduce a new particle identification method, ``Pratt-Walking'', 
and hybrid tracking algorithm for three-dimensional LPT.  
The former is able to differentiate between two overlapping particles to over 80\% overlap, is robust to noise, 
and outperforms the circular Hough Transform as well as local maximum methods.  
The hybrid algorithm is capable of tracking particles to separations approaching collision, 
tracks an order of magnitude more reliably than the 4 Frame: Best Estimate, 
and is able to measure collision rates with 95\% accuracy for tracking difficulty $\xi < 0.2$.  
Detecting collisions with high certainty imposes a greater restriction on the maximum tracking difficulty 
than does high-fidelity tracking of non-colliding trajectories; where maintaining $E_{track} < 0.1$ requires only $\xi < 0.7$, maintaining $C_{g} > 0.9$ requires $\xi < 0.2$. 
This work represents the first successful effort to develop a systematic tool for measuring collision rates of particles suspended in fluids.  

The hybrid tracking algorithm presented here is generalizable and modular.  
By relying on minimizing a penalty function that is the weighted sum of many elements, 
the method maintains flexibility for tracking distinct populations of particles.  
While here we use only three quantities in the penalty function (based on the positions of identified particles in four frames of data), 
this can be extended to any number of characteristics of the particles (size, shape, etc.) 
and an expectation for how they should evolve with time with marginal increase in computation time for each addition.  
It is necessary to construct this objective function as a robust metric for differentiating between particles as they approach contact in order to find collisions between trajectories.

\begin{acknowledgements}
We would like to thank Melanie Li Sing How and Lance Collins of Cornell University for providing DNS data of inertial coalescing particles in turbulence for testing the tracking algorithm presented here.
\end{acknowledgements}




\begin{thebibliography}{45}
\providecommand{\natexlab}[1]{#1}
\providecommand{\url}[1]{{#1}}
\providecommand{\urlprefix}{URL }
\expandafter\ifx\csname urlstyle\endcsname\relax
  \providecommand{\doi}[1]{DOI~\discretionary{}{}{}#1}\else
  \providecommand{\doi}{DOI~\discretionary{}{}{}\begingroup
  \urlstyle{rm}\Url}\fi
\providecommand{\eprint}[2][]{\url{#2}}

\bibitem[{Aliseda et~al.(2002)Aliseda, Cartellier, Hainaux, and
  Lasheras}]{Aliseda2002}
Aliseda A, Cartellier A, Hainaux F, Lasheras JC (2002) {Effect of preferential
  concentration on the settling velocity of heavy particles in homogeneous
  isotropic turbulence}. Journal of Fluid Mechanics 468:77--105,
  \doi{10.1017/S0022112002001593}

\bibitem[{Ayala et~al.(2008)Ayala, Rosa, Wang, and Grabowski}]{Ayala2008}
Ayala O, Rosa B, Wang LP, Grabowski WW (2008) {Effects of turbulence on the
  geometric collision rate of sedimenting droplets. Part 1. Results from direct
  numerical simulation}. New Journal of Physics 10,
  \doi{10.1088/1367-2630/10/7/075015}

\bibitem[{Bateson and Aliseda(2012)}]{Bateson2012}
Bateson CP, Aliseda A (2012) {Wind tunnel measurements of the preferential
  concentration of inertial droplets in homogeneous isotropic turbulence}.
  Experiments in Fluids 52(6):1373--1387, \doi{10.1007/s00348-011-1252-6}

\bibitem[{Betelin et~al.(2012)Betelin, Smirnov, Nikitin, Dushin, Kushnirenko,
  and Nerchenko}]{Betelin2012}
Betelin VB, Smirnov NN, Nikitin VF, Dushin VR, Kushnirenko AG, Nerchenko VA
  (2012) {Evaporation and ignition of droplets in combustion chambers modeling
  and simulation}. Acta Astronautica 70:23--35,
  \doi{10.1016/j.actaastro.2011.06.021},
  \urlprefix\url{http://dx.doi.org/10.1016/j.actaastro.2011.06.021}
  
\bibitem[{Bewley and Saw(2013)}]{Bewley2013}
 Bewley, Gregory P and Saw, Ewe-wei (2013) {Observation of the sling effect}
 New Journal of Physics 15,
 \doi{10.1088/1367-2630/15/8/083051}
 
\bibitem[{Bord{\'a}s et~al.(2013)Bord{\'a}s, Roloff, Th{\'e}venin, and
  Shaw}]{Bordas2013}
Bord{\'a}s R, Roloff C, Th{\'e}venin D, Shaw RA (2013) {Experimental
  determination of droplet collision rates in turbulence}. New Journal of
  Physics 15, \doi{10.1088/1367-2630/15/4/045010}

\bibitem[{Bourgeois and Lassalle(1971)}]{Bourgeois1971}
Bourgeois F, Lassalle JC (1971) {An extension of the Munkres algorithm for the
  assignment problem to rectangular matrices}. Communications of the ACM
  14(12):802--804, \doi{10.1145/362919.362945}

\bibitem[{Canny(1983)}]{Canny1983}
Canny J (1983) {Finding Edges and Lines in Images}. PhD thesis, Massachusetts
  Institute of Technology

\bibitem[{Chen et~al.(2006)Chen, Goto, and Vassilicos}]{Chen2006}
Chen L, Goto S, Vassilicos JC (2006) {Turbulent clustering of stagnation points
  and inertial particles}. Journal of Fluid Mechanics 553:143--154,
  \doi{10.1017/S0022112006009177}

\bibitem[{Deriche(1987)}]{Deriche1987}
Deriche R (1987) {Using Canny's criteria to derive a recursively implemented
  optimal edge detector}. International Journal of Computer Vision
  1(2):167--187, \doi{10.1007/BF00123164}

\bibitem[{Devenish et~al.(2012)Devenish, Bartello, Brenguier, Collins,
  Grabowski, Ijzermans, Malinowski, Reeks, Vassilicos, Wang, and
  Warhaft}]{Devenish2012}
Devenish BJ, Bartello P, Brenguier JL, Collins LR, Grabowski WW, Ijzermans RH,
  Malinowski SP, Reeks MW, Vassilicos JC, Wang LP, Warhaft Z (2012) {Droplet
  growth in warm turbulent clouds}. Quarterly Journal of the Royal
  Meteorological Society 138(667):1401--1429, \doi{10.1002/qj.1897}

\bibitem[{Duru et~al.(2007)Duru, Koch, and Cohen}]{Duru2007}
Duru P, Koch DL, Cohen C (2007) {Experimental study of turbulence-induced
  coalescence in aerosols}. International Journal of Multiphase Flow
  33(9):987--1005, \doi{10.1016/j.ijmultiphaseflow.2007.03.006}

\bibitem[{Good et~al.(2014)Good, Ireland, Bewley, Bodenschatz, Collins, and
  Warhaft}]{Good2014}
Good GH, Ireland PJ, Bewley GP, Bodenschatz E, Collins LR, Warhaft Z (2014)
  {Settling regimes of inertial particles in isotropic turbulence}. Journal of
  Fluid Mechanics 759:R3, \doi{10.1017/jfm.2014.602}

\bibitem[{Grabowski and Wang(2013)}]{Grabowski2013}
Grabowski WW, Wang Lp (2013) {Growth of Cloud Droplets in a Turbulent
  Environment}. Annu Rev Fluid Mech 45:293--326,
  \doi{10.1146/annurev-fluid-011212-140750}

\bibitem[{G{\"u}lan et~al.(2012)G{\"u}lan, L{\"u}thi, Holzner, Liberzon,
  Tsinober, and Kinzelbach}]{Gulan2012}
G{\"u}lan U, L{\"u}thi B, Holzner M, Liberzon A, Tsinober A, Kinzelbach W
  (2012) {Experimental study of aortic flow in the ascending aorta via Particle
  Tracking Velocimetry}. Experiments in Fluids 53(5):1469--1485,
  \doi{10.1007/s00348-012-1371-8}

\bibitem[{Holzner et~al.(2008)Holzner, Liberzon, Nikitin, L{\"u}thi,
  Kinzelbach, and Tsinober}]{Holzner2008}
Holzner M, Liberzon A, Nikitin N, L{\"u}thi B, Kinzelbach W, Tsinober A
  (2008) {A Lagrangian investigation of the small-scale features of turbulent
  entrainment through particle tracking and direct numerical simulation}.
  Journal of Fluid Mechanics 598:465--475, \doi{10.1017/S0022112008000141}

\bibitem[{Hoyer et~al.(2005)Hoyer, Holzner, L{\"u}thi, Guala, Liberzon, and
  Kinzelbach}]{Hoyer2005}
Hoyer K, Holzner M, L{\"u}thi B, Guala M, Liberzon A, Kinzelbach W (2005) {3D
  scanning particle tracking velocimetry}. Experiments in Fluids
  39(5):923--934, \doi{10.1007/s00348-005-0031-7}

\bibitem[{Ireland and Collins(2012)}]{Ireland2012}
Ireland PJ, Collins LR (2012) {Direct numerical simulation of inertial particle
  entrainment in a shearless mixing layer}. Journal of Fluid Mechanics
  704:301--332, \doi{10.1017/jfm.2012.241}

\bibitem[{Ireland et~al.(2013)Ireland, Vaithianathan, Sukheswalla, Ray, and
  Collins}]{Ireland2013}
Ireland PJ, Vaithianathan T, Sukheswalla PS, Ray B, Collins LR (2013)
  {Computers {\&} Fluids Highly parallel particle-laden flow solver for
  turbulence research}. Computers and Fluids 76:170--177,
  \doi{10.1016/j.compfluid.2013.01.020},
  \urlprefix\url{http://dx.doi.org/10.1016/j.compfluid.2013.01.020}

\bibitem[{Ireland et~al.(2016{\natexlab{a}})Ireland, Bragg, and
  Collins}]{Ireland2016}
Ireland PJ, Bragg AD, Collins LR (2016{\natexlab{a}}) {The effect of Reynolds
  number on inertial particle dynamics in isotropic turbulence . Part 1 .
  Simulations without gravitational effects}. Journal of Fluid Mechanics
  796:617--658, \doi{10.1017/jfm.2016.238}

\bibitem[{Ireland et~al.(2016{\natexlab{b}})Ireland, Bragg, and
  Collins}]{Ireland2016a}
Ireland PJ, Bragg AD, Collins LR (2016{\natexlab{b}}) {The effect of Reynolds
  number on inertial particle dynamics in isotropic turbulence . Part 2 .
  Simulations with gravitational effects}. Journal of Fluid Mechanics
  796:659--711, \doi{10.1017/jfm.2016.227}

\bibitem[{Kawanisi and Shiozaki(2008)}]{Kawanisi2008}
Kawanisi K, Shiozaki R (2008) {Turbulent effects on the settling velocity of
  suspended sediment}. Journal of Hydraulic Engineering 134(2):261--266,
  \doi{10.1061/(ASCE)0733-9429(2008)134}

\bibitem[{Kobayashi et~al.(1991)Kobayashi, White, and Abidi}]{Kobayashi1991}
Kobayashi H, White JL, Abidi AA (1991) {An Active Resistor Network for Gaussian
  Filtering of Images}. IEEE Journal of Solid-State Circuits 26(5):738--748,
  \doi{10.1109/4.78244}

\bibitem[{Lapp et~al.(2012)Lapp, Rohloff, Vollmer, and Hof}]{Lapp2012}
Lapp T, Rohloff M, Vollmer J, Hof B (2012) {Particle tracking for polydisperse
  sedimenting droplets in phase separation}. Experiments in Fluids
  52(5):1187--1200, \doi{10.1007/s00348-011-1243-7}
  
\bibitem[{{Li Sing How} and Collins(2020)}]{LiSingHow2020}
{Li Sing How} M, Collins LR (2020) {Direct Numerical Simulation of near contact 
  motion and coalescence of inertial droplets in turbulence}. Manuscript in preparation

\bibitem[{Maxey and Corrsin(1986)}]{Maxey}
Maxey, Corrsin (1986) {Gravitation Settling of Aerosol Particles in Randomly
  Oriented Cellular Flow Fields}. Journal of the Atmospheric Sciences
  43(11):1112--1134

\bibitem[{Obligado et~al.(2014)Obligado, Teitelbaum, and
  Cartellier}]{Obligado2014}
Obligado M, Teitelbaum T, Cartellier A (2014) {Preferential concentration of
  heavy particles in turbulence}. Journal of Turbulence 15(5):293--310,
  \doi{10.1080/14685248.2014.897710}

\bibitem[{Olivieri et~al.(2014)Olivieri, Picano, Sardina, Iudicone, and
  Brandt}]{Olivieri2014}
Olivieri S, Picano F, Sardina G, Iudicone D, Brandt L (2014) {The effect of the
  Basset history force on particle clustering in homogeneous and isotropic
  turbulence}. Physics of Fluids 26(4), \doi{10.1063/1.4871480}

\bibitem[{Ott and Mann(2000)}]{Ott2000}
Ott S, Mann J (2000) {An experimental investigation of the relative diffusion
  of particle pairs in three-dimensional turbulent flow}. Journal of Fluid
  Mechanics 422:207--223, \doi{10.1017/S0022112000001658}

\bibitem[{Ouellette et~al.(2006)Ouellette, Xu, and Bodenschatz}]{Ouellette2006}
Ouellette NT, Xu H, Bodenschatz E (2006) {A quantitative study of
  three-dimensional Lagrangian particle tracking algorithms}. Experiments in
  Fluids 40(2):301--313, \doi{10.1007/s00348-005-0068-7}

\bibitem[{Pratt(1987)}]{VaughanPratt1987}
Pratt V (1987) {Direct Least-Squares Fitting of Algebraic Surfaces}. Computer
  Graphics 21(4):145--152

\bibitem[{Qian and Law(1997)}]{Qian1997}
Qian J, Law CK (1997) {Regimes of coalescence and separation in droplet
  collision}. Journal of Fluid Mechanics 331:59--80,
  \doi{10.1017/S0022112096003722}

\bibitem[{Reade and Collins(2000)}]{Reade2000}
Reade WC, Collins LR (2000) {Effect of preferential concentration on turbulent
  collision rates}. Physics of Fluids 12(10):2530--2540,
  \doi{10.1063/1.1288515}

\bibitem[{Saffman and Turner(1956)}]{Saffman1956}
Saffman PG, Turner JS (1956) {On the collision of drops in turbulent clouds}.
  Journal of Fluid Mechanics 1(1):16--30, \doi{10.1017/S0022112056000020}

\bibitem[{Salazar et~al.(2008)Salazar, {De Jong}, Cao, Woodward, Meng, and
  Collins}]{Salazar2008}
Salazar JP, {De Jong} J, Cao L, Woodward SH, Meng H, Collins LR (2008)
  {Experimental and numerical investigation of inertial particle clustering in
  isotropic turbulence}. Journal of Fluid Mechanics 600:245--256,
  \doi{10.1017/S0022112008000372}

\bibitem[{Schanz et~al.(2016)Schanz, Gesemann, and Schr{\"o}der}]{Schanz2016}
Schanz D, Gesemann S, Schr{\"o}der A (2016) {Shake-The-Box: Lagrangian particle
  tracking at high particle image densities}. Experiments in Fluids
  57(5):1--27, \doi{10.1007/s00348-016-2157-1}

\bibitem[{Siewert et~al.(2014)Siewert, Kunnen, Meinke, and
  Schr{\"o}der}]{Siewert2014}
Siewert C, Kunnen RPJ, Meinke M, Schr{\"o}der W (2014) {Orientation statistics
  and settling velocity of ellipsoids in decaying turbulence}. Atmospheric
  Research 142:45--56, \doi{10.1016/j.atmosres.2013.08.011}

\bibitem[{Smoluchowski(1916)}]{Smoluchowski1916}
Smoluchowski M (1916) {Drei vortrage uber diffusion, brownsche bewegungund
  koagulation von kolloidteilchen}. International Journal of Research in
  Physical Chemistry and Chemical Physics 17:557--585

\bibitem[{Sobel(1968)}]{Sobel1968}
Sobel I (1968) {A 3 x 3 isotropic gradient operator for image processing}. In:
  Stanford Artificial Intelligence Project

\bibitem[{Sumbekova et~al.(2017)Sumbekova, Cartellier, Aliseda, and
  Bourgoin}]{Sumbekova2016}
Sumbekova S, Cartellier A, Aliseda A, Bourgoin M (2017) {Preferential
  Concentration of Inertial Sub-Kolmogorov Particles. The roles of mass loading
  of particles, St and Re}. Physical Review Fluids 2(2):1--18,
  \eprint{arXiv:1607.01256v1}

\bibitem[{Trujillo-Pino et~al.(2013)Trujillo-Pino, Krissian,
  Alem{\'a}n-Flores, and Santana-Cedr{\'e}s}]{Trujillo-Pino2013}
Trujillo-Pino A, Krissian K, Alem{\'a}n-Flores M, Santana-Cedr{\'e}s D
  (2013) {Accurate subpixel edge location based on partial area effect}. Image
  and Vision Computing 31(1):72--90, \doi{10.1016/j.imavis.2012.10.005},
  \urlprefix\url{http://dx.doi.org/10.1016/j.imavis.2012.10.005}

\bibitem[{Veenman et~al.(2003)Veenman, Reinders, and Backer}]{Veenman2003}
Veenman CJ, Reinders MJT, Backer E (2003) {Establishing motion correspondence
  using extended temporal scope}. Artificial Intelligence 145(1-2):227--243,
  \doi{10.1016/S0004-3702(02)00380-6}

\bibitem[{Virant and Dracos(1997)}]{Virant1997}
Virant M, Dracos T (1997) {3D PTV and its application on Lagrangian motion}.
  Measurement Science and Technology 8(12):1539--1552,
  \doi{10.1088/0957-0233/8/12/017}

\bibitem[{Wang et~al.(2000)Wang, Wexler, and Zhou}]{Wang2000}
Wang LP, Wexler AS, Zhou Y (2000) {Statistical mechanical description and
  modelling of turbulent collision of inertial particles}. Journal of Fluid
  Mechanics 415:117--153, \doi{10.1017/S0022112000008661}

\bibitem[{Xue et~al.(2008)Xue, Wang, and Grabowski}]{Xue2008}
Xue Y, Wang LP, Grabowski WW (2008) {Growth of Cloud Droplets by Turbulent
  Collision-Coalescence}. Journal of the Atmospheric Sciences 65(2):331--356,
  \doi{10.1175/2007jas2406.1}

\end{thebibliography}
\end{document}